\documentclass{llncs}
\usepackage{lncsedoc}
\usepackage[final]{graphicx}
\begin{document}
\title{A Lagrangian-Driven Cellular Automaton Supporting Quantum Field Theory}

\author{ Hans H. Diel}

\institute{Diel Software Beratung und Entwicklung, Seestr.102, 71067 Sindelfingen, Germany,
diel@netic.de}
\maketitle

\begin{abstract}
Models of areas of physics in terms of cellular automata have become increasingly popular. Cellular automata (CAs) support the modeling of systems with discrete state component values and enforce the comprehensive specification of the dynamic evolution of such systems.
Because many areas of physics can be described by starting with a specific Lagrangian, the idea to derive a cellular automaton directly from the Lagrangian (or similar construct, such as the Hamiltonian or action)
is not new. Previous work, however, indicated that the classical CA may not be a sufficient basis for the modeling of more advanced physics theories, such as quantum field theory. Specifically, the modeling of interactions in quantum field theory requires extensions and modifications of the classical CA. This paper describes a proposal for an extended cellular automaton that is suited for support of quantum field theory.

\end{abstract}


\section{Introduction}

Attempts to describe models of areas of physics in terms of cellular automata (CAs) reach back to 
E. Fredkin, 1990 \cite{Fredkin}, and S. Wolfram’s work  in 2002  \cite{Wolfram}. 
Some of these works applied cellular automata to areas of quantum theory (QT) (see \cite{Goessling},
\cite{Elze1}, \cite{Elze2}, \cite{thooft}). The relation of CAs to the Lagrangian or a related construct,
such as the Hamiltonian or an action, has been addressed in   \cite{Lee}, \cite{Elze1},  \cite{Elze2}.

Two items that make CAs especially attractive for the modeling of specific areas
of physics are the following:
\begin{enumerate}
\item CAs imply discreteness of the variables that constitute the system state. 
\\
Generally and traditionally, physics  theories do not assume discrete state variables. The discrete eigenstates of quantum theory represent only a special case of discreteness. Discreteness with other special aspects is addressed in \cite{thooft}, \cite{thooft2} and \cite{Dielfi}.  
\item CAs represent a functional model of the subject theory, i.e., a model that
specifies the dynamical evolution of the system in terms of state transitions.

\end{enumerate}
While these properties are desirable for special types of formulations of physics theories (e.g., \cite{thooft}, \cite{thooft2}, \cite{Dielfi}), they represent a challenge at the same time because existing standard formulations of areas of physics, in general, are well-suited neither for discreteness nor for functional descriptions.

After trying to formulate a Lagrangian-based and CA-based model of different
areas of physics, it turned out that the classical CA is not a sufficient basis for the
modeling of more advanced areas of physics (such as quantum field theory)
and that the Lagrangian provides much but not the complete content of the
physics theories. Thus, the objective of this paper is also the identification of
missing pieces and deficiencies and the provision of proposals to overcome these
deficiencies.

Section 2 contains a proposal for a CA called QFTCA, which is an extension of
the classical CA. In section 3, for some major areas of classical physics (e.g., Newtonian
mechanics, waves, fields, non-relativistic quantum mechanics), the Lagrangian is
transformed into a CA-based functional model. Quantum field theory (QFT), the
main subject of this paper, is addressed in sections 5 and 6. Section 6 describes
a CA-based functional model for interactions in QFT. The CAs are, as much
as possible, derived from the Lagrangian for the subject areas of physics. More
precisely, the CAs are derived from the equations of motion that are derived
from the Lagrangian. The transformation of the equation of motion into a CA,
which shows the dynamical state evolution in accordance with the equation
of motion, is relatively straightforward. However, the generation of a CA that
provides a more complete functional model of the subject area of physics requires
additional specifications that cannot be deduced solely from the equation of
motion. Therefore, the CA-based functional model of QFT interactions (section 6) is based
on the functional model of QFT interactions described in   \cite{Dielfi}.

The CA described in this paper has been accompanied by a computer model
QFTCA, which enables experimentation with alternative solutions and the generation
of the illustrations of examples contained in this paper. However, the
computer model is not the subject of the present paper. A rough overview of the
computer model is given in appendix A.  
 
\section{The Cellular Automaton}

\subsection{The classical Cellular Automaton}

The classical CA consists of a k-dimensional grid of cells. The state of the CA is given by the totality  of the states of the individual cells.

    state $ =  \{ s_{1}, ... , s_{n} \} $
\\
With traditional standard CAs, the cell states uniformly consist of the same state components

   $    s_{i} =  \{ s^{1}_{i}, ... , s^{j}_{i} \} $
\\
Typically, the number of state components, j, is 1, and the possible values are restricted to integer numbers.
The dynamical evolution of the CA is given by the "update-function", which computes the new 
state of a cell as a function of  its current state and the states of the neighbor cells.
\small{
\begin{verbatim}
Standard-CellularAutomaton(initial-state)  :=     // transition function
state = initial-state;
DO FOREVER {
   state = update-function(state, timestep);
   IF ( termination-state(state))  STOP;
}
\end{verbatim}
}
\normalsize
The full functionality and complexity of a particular cellular automaton  is concentrated in the update-function. As Wolfram (see \cite{Wolfram}) and others (see, for example, \cite{Ilachinski}) showed, a large variety of process types (e.g., stable, chaotic, pseudo-random, oscillating) can be achieved with relatively simple update-functions.

\subsection{The extended Cellular Automaton, QFTCA}

To enable support of the most advanced areas of physics (e.g., QT/QFT),
including support of multiple fields and particles, a cellular automaton QFTCA
is defined, which is an extension of the classical CA, made by embedding the CA into
an overall physics-based system structure. The embedding into the physics-based
system structure affects two aspects: the structure of the state of the system and
the functions for the dynamic evolution of the system.

\subsubsection{State structure}
The states of QFTCA cannot be expressed solely by the collection of cell states. In addition to the "local" information contained in the CA cells, non-local information about physical objects (e.g., waves, particles and fields)  have to be part of the system state. Position related (i.e., local) information of physical objects can be assigned to cell states. Other (non-local) information has to be kept in addition to the cell states.
\small{
\begin{verbatim}
System-state :=  
  CA-cells,    
  Particle-set,
  Field-set;
\end{verbatim}
}
\normalsize
The CA cells  \{$  s_{1}, ... , s_{n} $ \} represent the space. The particles and the fields have to be
mapped to the space, i.e., to the CA cells. The state components of an individual
cell depend on the particle or field to which the cell belongs. The set of state
components is thus not uniform for all cells. The possible values of the state
components are not restricted to integer values (however, a discrete value set
may be assumed with specific types of state components).
For the mapping of the overall system state to the CA cells, the following
hold true:
\begin{itemize}
\item For the QFTCA described in this paper, it is left open which state components are assigned to CA cells as opposed to state components that are kept with the physical objects. It has to be ensured that the cells (i.e., space
points) belonging to a physical object can be determined and vice versa; i.e., that
the physical objects occupying a CA cell can be determined.
\item A physical object may cover multiple cells.
\item A cell may be covered by multiple physical objects.
\item Time is \emph{not} a (fourth) dimension of QFTCA. Instead, the time derivatives such as  $  \partial \psi / \partial t, dx/dt $ and $ p^{\mu}$, are explicit parts of the system state.
\end{itemize}
The above-described state structure and timing considerations result in differing contents of CA cells and CA update-functions, which depend on the physical
objects assigned to the cells.

\subsubsection{Evolution of the system state} 
The structuring of the overall system state into  the CA cells and  the superimposed physical objects was also motivated by the differing update requirements of the physical objects.
\small{
\begin{verbatim}
QFTCA (initial-state)  :=  // transition function
state = initial-state;
DO FOREVER {
   state = global-update-function(state, timestep);
   IF ( termination-state(state))  STOP;
}

global-update-function(state, timestep)  :=    
DO PARALLEL {
  FOR ( all fields field[i] ) {
       field-state( field[i] ) = field-update-function(field[i], timestep);
  } 
  FOR ( all particles pw[k] ) {
        propertimestep = fx(pw[i] * timestep);
        pw[k] = pw-update-function(pw[k], propertimestep);
        IF ( interaction-occurred( pw[k], pw2 )  
            perform-interaction( pw[k], pw2 );	
   }
}
\end{verbatim}
}
\normalsize
The update-functions field-update-function(), pw-update-function() and perform-interaction() are derived from the Lagrangian and/or from the equation of motion of the pertaining field or particle type. Moreover, the CA has to be initialized,
partly by information derived from the Lagrangian and partly by additional
application-specific specifications. 

For the evolution of the overall system state, QFTCA assumes that the execution
proceeds in uniform global time steps. The update-function for the individual
particle/waves, however, has to proceed in proper time associated with
the particle/waves.

\subsection{From the Lagrangian to the extended Cellular Automaton}

\subsubsection{From the Lagrangian to the Equation of Motion}

The objective of a cellular automaton of an area of physics is to show the dynamical evolution of a physical system within the subject area.  Within physics the  dynamical evolution of a system is described by the \emph{equation(s) of motion}. Given a Lagrangian L, the equation of motion can be derived from the Lagrangian by using the Euler-Lagrange equation

(1) $ \frac{d}{dt} \frac{\delta L}{\delta \dot{x}}  - \frac{\delta L}{\delta x}  = 0 $.
\\
Example 1: The Lagrangian for classical mechanics: L = V - T with 
$ V=V(x), T = \frac{1}{2} m \dot{x}^{2} $  (see, for example \cite{McMahon}, page 24) 
The Euler-Lagrange equation leads to
 
(a) $ \frac{d}{dt} \frac{\delta L}{\delta \dot{x}} =  \frac{d}{dt}  \frac{\delta( \frac{1}{2} m \dot{x}^{2} - V(x))  }{\delta \dot{x}} =  \frac{d}{dt}  \frac{\delta( \frac{1}{2} m \dot{x}^{2})  }{\delta \dot{x}}- \frac{d}{dt}  \frac{\delta( V(x))  }{\delta \dot{x}} =  \frac{d}{dt} ( m \dot{x}) = m \ddot{x} $
 
(b) $  \frac{\delta L}{\delta x}  =   \frac{\delta( \frac{1}{2} m \dot{x}^{2} - V(x))}{\delta x}  =   \frac{\delta( \frac{1}{2} m \dot{x}^{2})}{\delta x} -  \frac{\delta( V(x)}{\delta x}= 0 -  \frac{\delta V}{\delta x} $.
\\
Equating (a) to (b) gives 

 $     m \ddot{x} =  \frac{\delta V}{\delta x} $
\\
Assuming V(x) caused by a force F, such that $ F = -  \frac{\delta V}{\delta x} $ gives 

 (2) $  F = m  \ddot{x} $
\\
With this example, the objective of the CA would be to show the dynamical evolution of x, the position of the moving particle.
Thus, the CA update-function has to provide the dynamical update of the system state, in particular the update of x. Equation (2) enables the computation of $  \ddot{x} $, provided the other variables appearing within (2) are known. From $  \ddot{x} $,  the updated value of x can be computed provided the actual values of $  \dot{x} $ and x are known. 

The above described process for the derivation of the CA update-function from the Lagrangian or the equation of motion is not considered to be part of the
CA execution. It is rather part of the process that generates the CA update-function
from a given Lagrangian (or equation of motion). The computer model
QFTCA
 (see \cite{QFTCA}) supports in addition to the executing CA the CA generation.

\subsubsection{From the Equation of Motion to the CA Update-Function} The CA update-function is invoked repetitively from the CA to update the system state in constant time intervals. Two cases of system state update can be distinguished:
\begin{enumerate}
\item Update of the particle state, primarily the particle position,
\item Update of fields, primarily the field state variable $ \psi(x, t) $ assigned to the CA
cells.
\end{enumerate}
For both cases, the equation of motion has to be interpreted. The particle positions
can be updated if the equation of motion delivers 
 $  \ddot{x} $ (or $ \dot{x}$ or x ) (see sections 3.1 and 3.2  for more details). The field state can be updated if the equation of motion delivers 
 $  \frac{d \psi^{2}}{d^{2}t}$ (or  $ \frac{d \psi }{dt} $) (see sections 3.3 and 3.5 for more details).

\subsubsection{Initialization of the CA}

The Lagrangian for a specific area of physics is
typically a very compact and very abstract equation. To obtain a runnable automaton
that shows the dynamical evolution of a system, an initial state with
initial values for all state components is required. As will be shown in the following
sections, there exist special constraints for the "valid" initial states for the
various physics areas and their Lagrangians. These constraints cannot be derived
from the Lagrangian but are usually hidden in additional laws of physics.

\section{Classical Physics}
\subsection{Classical (Newtonian) Mechaniscs}

\begin{figure}[ht]
\center{\includegraphics*[scale=0.5] {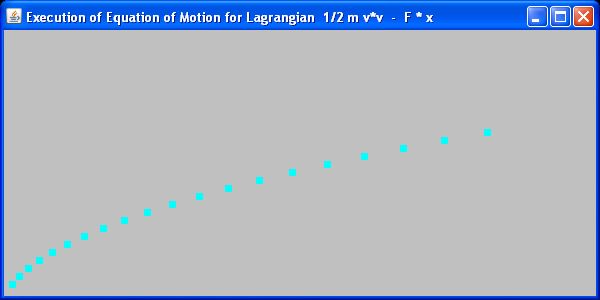} }
\caption{CA run for $ L = \frac{1}{2} m \dot{x}^{2} - V(x) $. \label{fig1}}
\end{figure}
The basic Lagrangian mechanism, as originally introduced by Lagrange, defines  

$  L = T - V $
\\
where T = kinetic energy, V = potential energy. As described in section 2.3, the Euler-Lagrange equation allows the generation of the equation of motion   

 $     m \ddot{x} =  \frac{\delta V}{\delta x} $.
\\
To execute the CA, initial settings are required for all entities (and certain derivatives of these entities) appearing in the equation of motion. Setting the initial (and constant) value for $  \frac{\delta V}{\delta x} = F $, i.e., assuming a constant power F, leads to the execution of  

 $  F = m  \ddot{x} $ 
\\
for the specified initial values. An example is illustrated in Fig. 1.
\\
Instead of specifying  $  \frac{\delta V}{\delta x} = F $ it is also possible to initialize the potential V(x) explicitly by a specific spatial distribution and initial values for all CA cells
$   \{ s_{1}, ... , s_{n} \} $. This enables more general multi-dimensional modeling.
  


\subsection{Harmonic Oscillator}
A particle of mass m undergoing simple harmonic motion is described by the Lagrangian

$  L = T - V =  \frac{1}{2} m \dot{x}^{2} - \frac{1}{2} k x^{2} $ (see, for example,  \cite{McMahon}, page 25). 
\\
The Euler-Lagrange equation leads to
 
(a) $ \frac{d}{dt} \frac{\delta L}{\delta \dot{x}} =  \frac{d}{dt}  \frac{\delta( \frac{1}{2} m \dot{x}^{2} - V(x))  }{\delta \dot{x}} =  \frac{d}{dt}  \frac{\delta( \frac{1}{2} m \dot{x}^{2})  }{\delta \dot{x}}- \frac{d}{dt}  \frac{\delta( \frac{1}{2} k x^{2} )  }{\delta \dot{x}} =  \frac{d}{dt} ( m \dot{x}) = m \ddot{x} $
 
(b) $  \frac{\delta L}{\delta x}  =   \frac{\delta( \frac{1}{2} m \dot{x}^{2} - V(x))}{\delta x}  =   \frac{\delta( \frac{1}{2} m \dot{x}^{2})}{\delta x} -  \frac{\delta( V(x)}{\delta x}= 0 -  \frac{\delta V}{\delta x} = - kx $
\\
Equating (a) to (b) gives the equation of motion

 $     m \ddot{x} =  -k x $.
\\
%
\begin{figure}[ht]
\center{\includegraphics*[scale=0.5] {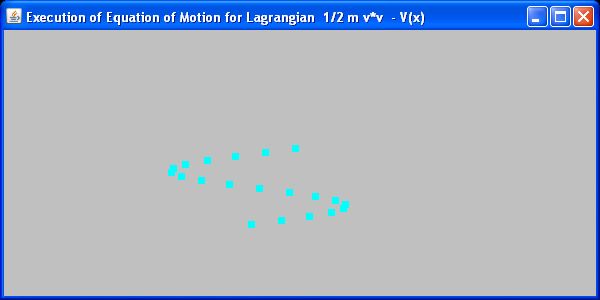} }
\caption{CA run  for harmonic oscillator $ L = \frac{1}{2} m \dot{x}^{2} - \frac{1}{2} k x^{2} $ . \label{fig2}}
\end{figure}

With \cite{Strassler} the equation of motion for the harmonic oscillator is

$ d^{2}z/dt^{2} = - K/M (z -  z0)   $  $ \Rightarrow  d^{2}x/dt^{2} = - K/M  x + K/M \cdot x0    $

 $ \Rightarrow  d^{2}x / dt^{2} =   - K/M \cdot x  $  (in case  x0 = 0).

Fig. 2 shows a snapshot of running QFTCA with the Lagrangian  for the harmonic oscillator $  L =  \frac{1}{2} m \dot{x}^{2} - \frac{1}{2} k x^{2} $ . 

\subsection{Waves}

The standard formulation of the "wave equation", i.e., the equation of motion for waves, is (see, for example \cite{Chun})

(3)  $ ( \frac{1}{v^{2}} \frac{d^{2}}{dt^{2}} - \frac{d^{2}}{dx^{2}}) \psi (x,t) = 0 $.
\\
The three-dimensional version is

(4)	$ (\frac{1}{v^{2}} \frac{  \partial^{2} }{ \partial t^{2}} - \bigtriangledown^{2} ) \psi (x,t) = 0 $.
\\
Depending on the particular context, the equation may be varied or extended
by setting the right hand side not equal to 0. For example,  in \cite{Strassler} the equation of motion for class 1 waves is

(5) $ d^{2}\psi/dt^{2} - c_{w}^{2}  d^{2}\psi/dx^{2} = – (2 \pi \nu)^{2} (\psi - \psi_{0})   $.
\\
To obtain the CA update-function, the equation of motion has to be transformed into a sequence of computation steps, including the replacement of the differential operations by discrete "$ \Delta $-units". The following computation steps for the CA  are derived from equation (3) for the state transition $ \psi(x_{i},t_{j})  \rightarrow \psi(x_{i},t_{j+1}) $
\begin{enumerate}
\item $ t_{j+1} = t_{j} + \Delta t $
\item $  \Delta \psi dx = ( \psi(x_{i+1},t_{j}) - \psi(x_{i-1},t_{j}) ) / 2  \Delta x $
\item $ \Delta^{ 2} \psi dx  = ( \psi(x_{i+1},t_{j}) - 2  \psi(x_{i},t_{j}) + \psi(x_{i-1},t_{j})) / \Delta x /  \Delta x $
\item equation of motion derived from (3): $ \Delta^{ 2} \psi dt =  v^{2}  \Delta^{ 2} \psi dx  $
\item $ \Delta^{ 2} \psi dt =  ( \psi(x_{i},t_{j+1}) - 2  \psi(x_{i},t_{j}) + \psi(x_{i},t_{j-1})) / \Delta t /  \Delta t 
\rightarrow  $
\\
$  \psi(x_{i},t_{j+1}) = \Delta^{ 2} \psi dt \cdot \Delta t \cdot  \Delta t +  2  \psi(x_{i},t_{j}) - \psi(x_{i},t_{j-1}) $
\end{enumerate}  
Here and subsequently, the following naming conventions are used for going from the differential units to $ \Delta$ -units: $ d \psi/dt \rightarrow \Delta \psi dt $, $ d \psi/dx \rightarrow \Delta \psi dx $,   $ d^{2} \psi / dx^{2}  \rightarrow  \Delta^{ 2} \psi dx $, $ d^{2} \psi / dt^{2}  \rightarrow  \Delta^{ 2} \psi dt $.

The above sequence of CA computation steps can be generalized and is applicable
to arbitrary Lagrangians, except for the Lagrangian-specific step 4. The
computation requires an equation of motion that delivers 
 $ \frac{d^{2} \psi}{dt^{2}} $ and an initial system state that contains initial values for $  \psi (x,t_{0}) $ and $  \psi (x,t_{-1}) $ (or, alternatively, $ \Delta \psi dt $) for the relevant positions x.
The CA execution has to start with an initial state that represents a specific physical situation.

Notice that the Lagrangian and thus the system state of the CA does not include variables that are usually included in physics theories of waves such as frequency, amplitude, wavelength and energy. 
These variables are implied by the initial state of the wave and explicitly supported  by the CA only if they appear in the equation of motion. Such hidden variables may also appear in
the constraints for the initial state mentioned in 2.3.

Fig.3 shows a snapshot from the dynamical evolution of a wave according to the equation of motion 
$ ( \frac{1}{v^{2}} \frac{d^{2}}{dt^{2}} - \frac{d^{2}}{dx^{2}}) \psi (x,t) = 0 $ and a suitable initial state.
\begin{figure}[ht]
\center{\includegraphics*[scale=0.5] {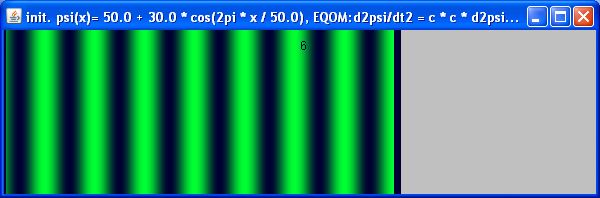} }
\caption{CA run for wave $ ( \frac{1}{v^{2}} \frac{d^{2}}{dt^{2}} - \frac{d^{2}}{dx^{2}}) \psi = 0 $  \label{fig3}}
\end{figure}

\subsection{Fields}

Fields are based on waves.
As with waves (section 3.3), instead of the position variables  $ x $, $ \dot{x}$  and $ \ddot{x} $ of the particles, the  Lagrangian
and related system state contain the entities  $ \psi(x,t) $ assigned to spacetime points (x,t). 
The Lagrangian (and the equation of motion) typically depends on $ \psi(x,t) $ and the derivatives  $ \partial \psi /  \partial t, 
 \partial \psi /  \partial x,
 \partial^{2} \psi / \partial t^{2},   \partial^{2} \psi / \partial x^{2} $. 

The equation of motion (which is also called the field equation)
is obtained by using the EL equation (1) or by minimizing the action.
Depending on the type of field considered, $ \psi (x,t) $ may be complex and may represent a vector, matrix, or tensor. 
The Lagrangian may also describe multiple fields, which means that multiple field variables such as $ \psi_{1}, \psi_{2}  $ or 
$ \varphi_{1}, \varphi_{2}  $, ... may be required.
\\
For fields, the \emph{Lagrangian density} $  \mathcal{L} $ is defined such that 

$  L   = \int \mathcal{L} d^{3} x $  (see \cite{Zee}, pages 16, 17).
\\
In \cite{Bogoljubov} (page 26), the following requirements for Lagrangian density are listed:
\begin{enumerate}
\item The Lagrangian density must be a function of the dynamical variables only, i.e., a function of the fields  $ \psi(x,t) $ and their derivatives.
\item It must not contain any explicit dependencies on the coordinates x.
\item The dependency must be local, i.e., it must depend on $ \psi(x,t) $ and partial derivatives of $ \psi(x,t) $. To obtain an equation of motion with less than third order derivatives, the derivatives of $ \psi(x,t) $ should not be higher than first order.   
\item The Lagrangian density must be a real number.
\item A number of symmetry requirements hold. For relativistic theories, the most important symmetry requirement  is the requirement for Poincare invariance.
\end{enumerate}
These requirements apply primarily to the Lagrangian of the respective physics theories. For the mapping of the Lagrangian to a cellular automaton the requirements represent an alleviation rather than a problem because they reduce the generality.

\subsection{Non-relativistic Quantum Mechanics}

In non-relativistic quantum mechanics (QM), the equation of motion is  the Schr\"odinger equation. For a particle moving in one direction, it is 

(6)  $ -  \frac{\hbar^{2}}{2m}  \frac{\partial^{2}\psi(x,t)}{\partial x^{2}} + V(x,t)  \psi(x,t) = i \hbar  \frac{\partial \psi(x,t)}{\partial t} $.
\\
The update-function of the CA continuously determines $   \frac{\partial \psi}{\partial t} $ as to

 (7) $  \frac{\partial \psi}{\partial t} =   \frac{V  \psi }{ i \hbar  } -  \frac{\hbar}{2mi}  \frac{\partial^{2}\psi}{\partial x^{2}} $.
\\
Because the equation of motion delivers $  d \psi/dt $ instead of  $ d^{2} \psi / dt^{2} $ described in section 3.3 for waves, the computation sequence shown in section 3.3 for the state transition $ \psi(x_{i},t_{j})  \rightarrow \psi(x_{i},t_{j+1}) $ is slightly varied to become 
\begin{enumerate}
\item $ t_{j+1} = t_{j} + \Delta t $
\item $  \Delta \psi dx = ( \psi(x_{i+1},t_{j}) - \psi(x_{i-1},t_{j}) ) / 2  \Delta x $
\item $ \Delta^{ 2} \psi dx  = ( \psi(x_{i+1},t_{j}) - 2  \psi(x_{i},t_{j}) + \psi(x_{i-1},t_{j})) / \Delta x /  \Delta x $
\item equation of motion derived from (7): $  \Delta \psi dt =   \frac{V  \psi }{ i \hbar  } -  \frac{\hbar}{2mi} \Delta^{ 2} \psi dx  $
\item $ \Delta \psi dt =  \Delta \psi dt + \Delta^{ 2} \psi dt \cdot  \Delta t$
\item $ \psi(x_{i},t_{j+1}) =  \psi(x_{i},t_{j}) + \Delta \psi dt \cdot  \Delta t $
\end{enumerate} 
The naming conventions are those described in section 3.3.
With quantum theory $ \psi(x,t) $ is a complex scalar number. 

Many laws of quantum mechanics that are usually described in textbooks
for QM cannot directly be deduced from the Schr\"odinger equation (or the related
Lagrangian). Major examples are the uncertainty relation, the rules explaining
the double slit experiment and the measurement problem. As a consequence, it is
not possible to derive a CA that supports full QM purely from the Schr\"odinger
equation. The author claims that a major part of the required additional specifications can only be provided by a functional model of QT that includes QT interactions.
In \cite{Dielfi}, such a functional model of QT interactions is proposed and is the basis for section 5.

\section{Quantum Fields and Particles}

Compared to (non-relativistic) quantum mechanics, quantum field theory is a relativistic theory of quantum fields and particles  and interactions among fields and particles. Particles are largely treated like fields. In the literature, for
example,  \cite{Strassler}, particles are called "quanta of fields". Various field types with differing Lagrangians  are distinguished. 
For the purpose of this paper, it is sufficient to consider the field(s)  of quantum electrodynamics (QED). 
In  \cite{McMahon} (page 170), the Lagrangian of QED is given by 
 
 $ L_{QED} = L_{EM} + L_{Dirac} + L_{int} $
\\
with

$ L_{EM} =  - \frac{1}{4} F_{\mu \nu} F^{\mu \nu} $

$  L_{Dirac} = i \bar{ \psi } \gamma^{\mu} \partial_{\mu}  \psi  - m \bar{ \psi }  \psi  $

$  L_{int} = -q \bar{ \psi } \gamma^{\mu} \psi  A_{\mu}  $.
\\
Thus

(8) $ L_{QED} = - \frac{1}{4} F_{\mu \nu} F^{\mu \nu} +  i \bar{ \psi } \gamma^{\mu} \partial_{\mu}  \psi  - m \bar{ \psi }  \psi  - q \bar{ \psi } \gamma^{\mu} \psi  A_{\mu} $
\\
The equations refer to two types of fields: the electromagnetic  field and the Dirac field. Leaving aside interactions between  fields and particles (which will be addressed in section 5), it is sufficient to consider the equations of motion for the  individual field types and particle types separately. The equations of motion can be derived directly from the respective Lagrangian parts. For example, the equation of motion (called the field equation) for the Dirac field, the so-called Dirac
equation, is

(9)  $ i \hbar \gamma^{\nu} \frac {\partial_{\psi} } {\partial x^{\nu} } - m c \psi = 0 $.
\\
Thus, a QFTCA that supports QFT, or a specific part of QFT, such as QED,
has to have different update-functions depending on the different field types
and particle types to be supported. Additionally, the system state encompassing physics
objects and cell states depends on the field types and particle types. 

CA applications with multiple fields and/or particles may be viewed as the union of multiple independent field-specific CAs (as long as interactions are disregarded).

\section{Interacting Quantum Fields and Particles}

This section considers interactions among particles and interactions between
particles and fields. The considerations focus on interactions between two "in"
particles resulting in one or two "out" particles. Only interactions are considered,
which, in QFT, are dealt with by scattering matrices and Feynman diagrams (in \cite{Dielfi} and in the following such interactions are called "QFT interactions"). 

Many details of QFT concerning interacting particles/fields can be derived from the
interaction part of the Lagrangian. For QED, the interaction part of the Lagrangian in equation (8) is

(10) $  L_{int} = -q \bar{ \psi } \gamma^{\mu} \psi  A_{\mu}  $
\\
In contrast to the areas described in the preceding sections, a functional model of QFT interactions  cannot solely be derived from the Lagrangian. Additional assumptions have to be specified to obtain a complete functional model for QFT interactions. In \cite{Dielfi}, such a functional model of QFT interactions is described and is taken as the basis for  the CA-based functional model described in the present section. 
The following is a summary of the functional model of QFT interactions as described in \cite{Dielfi}, mapped to the structure and concepts of QFTCA and partly providing more details in terms of QFTCA. 

\subsubsection{Refinement of the system state  structure}
To support a CA-based functional model of QFT interactions the system structure described in section 2.2 has to be refined. 

QFT provides an extensive framework for the more detailed evaluation of probability
amplitudes for various types of interactions. 
This framework is based on Feynman diagrams and related "Feynman rules". Rules have been established for the derivation of the possible Feynman diagrams from a given Lagrangian.

With QFT (see, for example,  \cite{Weinberg}), an interaction (e.g., scattering) is described by the scattering  matrix
 (S-matrix), which assigns a probability amplitude  $ S_{\beta \alpha}$ to  the transition of a given "in"  state $ \Psi_{\alpha} $ to an "out" state $ \Psi_{\beta} $.
\\
 $ S_{\beta \alpha} = ( \Psi_{\beta}, \Psi_{\alpha} )  $
\\
The "in" and "out" states are specified by their state components  
\\
  $  \Psi_{\alpha} = \{inparticle1( p1,\sigma1,n1), inparticle2( p2,\sigma2,n2) \}$; 
\\
  $ \Psi_{\beta} =  \{outparticle1( p1',\sigma1',n1'), outparticle2( p2',\sigma2',n2') \} $.
\\
Here, $ p_{i} $ represents the momentum, $ \sigma_{i} $ is the spin, and $ n_{i} $ is the particle type. For a more complete specification of a state, the position x and the time t are also part of the state.
\footnote{An explanation of the reasons why position and time are  (usually) not parameters of the QFT scattering matrix computation is outside the scope of this paper.}
For the computation of a QFT scattering matrix element, a concrete in-state $  \Psi_{\alpha} $ and a concrete 
out-state $ \Psi_{\beta} $ are typically assumed.
For a functional model of QFT, however, the dynamical evolution of the system state does not end with a concrete and definite interaction result but may continue with a multitude of alternative interaction results. Only a measurement (by use of other QFT interactions) reduces the multitude of possible scattering results to a single definite result.
Thus, for a functional model, the complete interaction result consists of  all the possible out-state combinations with associated probability amplitudes.
\\
 $  outstate = \{ $
\\
 $ \{ amplitude_{1}, outparticle1( t, x1_{1}, p1_{1},\sigma1_{1},n1), outparticle2( t, x2_{1}, p2_{1},\sigma2_{1},n2) \} $,
\\
     $ \{ amplitude_{2}, outparticle1(  t, x1_{2}, p1_{1},\sigma1_{2},n1), outparticle2( t, x2_{2}, p2_{2},\sigma2_{2},n2) \} $,
\\ ...
\\
  $ \{ amplitude_{i}, outparticle1( t, x1_{i}, p1_{i},\sigma1_{i},n1), outparticle2( t, x2_{i}, p2_{i},\sigma2_{i},n2) \} $,
\\ ...
\\ \}
\\
This interaction result suggests a huge multitude of state combinations (with associated probability amplitudes). 
The multitude is reduced due to two points:
\begin{enumerate}
\item Only specific combinations of outparticle1() and outparticle2() are possible (for example, because of energy and momentum conservation).
\\
(outstate is not a product state of outparticle1-state and outparticle2-state).
\item Only a discrete set of alternative out states (called "paths") is assumed with the CA-based functional model.
\end{enumerate}
In  \cite{Dielfi}, the above state structure, which represents the interaction results, is called a particle/wave-collection (pw-collection). A tabulated representation of a pw-collection is shown in Table 1.
\begin{table}
\caption{\label{label}Structure of a pw-collection consisting of two particle/waves pw1 and pw2 and N paths.}
\begin{tabular} { | c | c | c  | c | }
\hline
paths & pw1-state & pw2-state  & amplitude  \\

\hline

path-1	  &   pw1-state$_{1}  $    &  pw2-state$_{1} $  & ampl-1  \\

path-2	 &   pw1-state$_{2}  $    &  pw2-state$_{2} $   & ampl-2  \\

...	            & ...         & ...    & ...  \\

path-N	 &  pw1-state$_{N} $    &  pw2-state$_{N} $  & ampl-N  \\
\hline
\end{tabular}
\end{table} 

\subsubsection{Occurrence of interaction}

In terms of wave equations (i.e.,  the equations of motion for waves) (see \cite{Strassler}),  an interaction between two waves 
$ \psi_{1} $ and $ \psi_{2} $ resulting in a third wave $ \psi_{3} $ is described by an equation of motion in which the product of waves  $ \psi_{1} $ and $ \psi_{2} $ is related to $ \psi_{3} $ as, for example, 

 $ d^{2}\psi_{3}/dt^{2} - c^{2} d^{2}\psi_{3}/dx^{2} =  a^{2} \psi_{3} + b \cdot \psi_{1} \psi_{2} $ 
\\
(Typically, similar equations exist defining the interactions between $ \psi_{1} $ and $ \psi_{3} $ and between $ \psi_{2} $ and $ \psi_{3} $.)
\\
An interaction occurs if, for a position $ x_{0} $ the product $  \psi_{1}(x_{0},t) \cdot \psi_{2}(x_{0},t ) $ becomes non-zero, which means that both $ \psi_{1} $ and $ \psi_{2} $ have to be non-zero. 

In terms of a CA-based model, such as QFTCA, an interaction occurs if, for
a cell $ c_{0} $, two fields $ \psi_{1} $ and $ \psi_{2} $ have a non-zero value.
This appears to be trivial, but it does
not specify what occurs if, at a given CA-update cycle, multiple cells satisfy this
condition. In line with the functional model of QFT interactions described in 
 \cite{Dielfi}, the following mechanism is assumed with QFTCA:
\begin{itemize}
\item Only one single (discrete) position (i.e, CA cell) may cause an interaction and the interaction results depend on the state attributes associated with this position. 
\item In general, a QFT interaction results in the creation of new "out" particles/waves (which may be of the same types as the "in" particles/waves). This may be viewed as a "collapse of the ingoing particle/wave" and a reduction of the "in" particles/waves to the definite attribute values associated with the interaction position.
\item If multiple positions (i.e., cells) satisfy the condition for the occurrence of an interaction, the single interaction position is randomly (i.e., non-deterministically) selected as a function of the probability amplitudes of the candidate cells/positions.  
\end{itemize}
This mechanism enables a modeling of  QFT interactions (e.g., scatterings). It also 
offers a model of QT measurement in which measurements are realized by
"normal" QFT interactions (see also section 6.5).
After an interaction has been initiated (and the occurrence of multiple interactions
of the same type has been blocked), the interaction process is performed.
In \cite{Dielfi}, the interaction process is subdivided into the three process steps. These process steps have to be performed as part of the CA update-function:

\small{
\begin{verbatim}
perform-interaction ::= {
  Step0: Occurrence of interaction;
  Step1: Formation of interaction-object;
  Step2: Formation and processing of interaction channels;
  Step3: Generation of "out" particle/waves.
}
\end{verbatim}
}
\normalsize

\subsubsection{Formation of interaction object}

At the beginning of a QFT interaction, the information from the interacting particles/waves is merged into an "interaction object". At the end of the interaction, the interaction object is replaced by the  the "out" particles/waves. 

The functional model assumes that interaction objects, similar to virtual
particles, have a limited life-time before they decay into the "out"  particles/waves that
are the result of the interaction. In contrast to virtual particles, 
the interaction object does not correspond to a single real particle, but contains the information from both interacting
particles.

\subsubsection{Formation and processing of interaction channels}

Starting with the interaction object, the CA update-function continues with the determination of the probability amplitudes for the possible results of the QFT interaction. In standard QFT the computation of the probability
amplitudes  is based on
Feynman diagrams, rules (called Feynman rules) and the assumption of "virtual
particles". The CA update-function (of course) cannot use Feynman diagrams
for the computation of state transitions toward the interaction results. Instead,
the functional model of QFT interactions assumes  interaction channels.

Interaction-channels are generated according to rules that are equivalent to the Feynman rules for the generation of Feynman diagrams. Like the Feynman rules, the interaction-channel generation rules are related to the interaction part of the Lagrangian (or the equation of motion).
As an example, consider the interaction part of the QED Lagrangian, which in equation  (10) is given by 

  $  L_{int} = -q \bar{ \psi } \gamma^{\mu} \psi  A_{\mu}  $.
\\
Here, $ \psi $ represents the electron field, $ \bar{ \psi } $ is the positron field, and  $ A  $ is the photon (i.e., electromagnetic) field. Thus, the equation applies to interactions between electrons $ e^{-} $, positrons $ e^{+} $, and photons $ \gamma $.   
The interaction is expressed in terms of creation and annihilation operators by

(12)  $ H_{W}(x) = -eN \{ ( \bar{\psi^{+} } +  \bar{\psi^{-} }) ( \not A^{+} + \not A^{-}) ( \psi^{+} -  \psi^{-}) \}_{x} $
\\
where $ \bar{ \psi^{+}}, \bar{ \psi^{-}}, \not A^{+}, \not A^{-},  \psi^{+},  \psi^{-} $ are creation and annihilation operators.
This includes four variants of photon absorption, $ e^{-} + \gamma  \rightarrow e^{-} $, $ e^{+} + \gamma  \rightarrow e^{+} $, $   e^{-} + e^{+} + \gamma $,  $  \gamma  \rightarrow e^{-} + e^{+} $ and four variants of photon emission, $ e^{-} \rightarrow e^{-} + \gamma $, $ e^{+} \rightarrow e^{+} + \gamma $, $ e^{-} + e^{+}  \rightarrow  \gamma $,   $  \rightarrow e^{-} + e^{+} + \gamma $ ( see \cite{Mandl}, page 111). 

Each interaction-channel starts with both "in" particles/waves and ends with "out" particles/waves. Alternative interaction-channels may differ in the set of "out" particles/waves and/or in the sub-channels between the "in" and "out" particles/waves.

A specific interaction-channel is formed by the combination of the two operators
split() and combine().

 split(a) $ \rightarrow $  (b,c)  means   that split(a) results in b and c;

combine(a,b)$  \rightarrow $  (c)  means that combine(a,b) results in c.
\\
For example, starting with two interacting particles/waves pw1 and pw2, the interaction-channel specified by
\\
  (pw1,pw2): split(pw1) $  \rightarrow $ (a,b);  combine( a, pw2 ) $ \rightarrow $ ( c)   
\\ would result in "out" particles/waves b and c.

The split() and combine() operators are analogous (although not equal)  to the creation and annihilation operators of QFT. 
Derived from QFT, the following rules are established for the functional model of interactions:
\begin{itemize}
\item Rule1: An interaction always starts with two "in" particles/waves and ends with two "out" particles/waves.
\item Rule2: An interaction-channel always contains one combine() and one split() (in arbitrary sequence).
\footnote{In QFT, deviations from these rules can be found; however, they will not be addressed here.} 
\end{itemize}
Given the above rules, there are five possible interaction-channels that can be constructed from the two interacting particles/waves pw1 and pw2:
\begin{enumerate}
\item combine( pw1, pw2 )   $  \rightarrow $ (a)  split(a)  $  \rightarrow $ (b,c)
\item split(pw1)  $  \rightarrow $ (a,b) combine( a, pw2 ) $  \rightarrow $ (c)
\item split(pw1)  $  \rightarrow $  (a,b) combine( b, pw2 ) $  \rightarrow $ (c)
\item split(pw2)  $  \rightarrow $  (a,b) combine( pw1, a ) $  \rightarrow $ (c)
\item split(pw2)  $  \rightarrow $  (a,b) combine( pw1, b ) $  \rightarrow $ (c)
\end{enumerate}
All these interaction-channels end with two particles/waves (a, c) or (b, c). Depending on the type of particles/waves involved in the interaction, QFT supports only specific types of splits and combines. The rules that define what combinations of particle types may be subject to combine(p1, p2) and what the resulting particle types of split(p1) and combine(p1, p2) can be are equivalent to the rules regarding the possible vertices of Feynman diagrams, as described in numerous textbooks on QFT 
 (see for example, \cite{Ryder}, \cite{Griffiths},  \cite{Mandl}).
 For QED, for example, one of the rules is
 split(  $ \gamma) \rightarrow  ( e^{-},  e^{+}) $;   combine($ e^{-}, \gamma) \rightarrow ( e^{-}) $ .
\\
Sometimes QFT rules allow for multiple results for split() and combine(). For example, split(photon) may result in (electron, positron), (muon, antimuon), or (tauon, antitauon). Finally, with specific "in" particle/wave combinations, it may turn out that some of the possible interaction-channels may be considered equivalent, and therefore only one of them has to be included.
\footnote{The rules regarding when interaction-channels may be considered equivalent are defined with QFT (in terms of equivalent Feynman diagrams) and are not addressed here.}
Given the QFT rules, typically only one or two of the above mentioned five alternative interaction-channels are possible for a specific "in" particle combination.

The operator split(a) $\rightarrow $ (b, c) is a non-bijective function insofar, as there are many alternatives with respect to the attributes (e.g., momentum and spin) of the resulting (b, c). Rather than selecting a particular specific result, QFT (and the functional model of QT interactions) requires that a multitude of possible results are generated with differing probability amplitudes assigned. 
\footnote{The functional model of QT interactions assumes a certain granularity for a multitude of possible results.}
Thus,   split(a) $ \rightarrow $ (b, c) results in a two-fold splitting and may be expressed as

split(a) $ \rightarrow  ((b_{1}, c_{1}),  (b_{2}, c_{2}), ... (b_{n}, c_{n})) $. 
\\
For the interaction object this means that a multitude of paths representing $  ((b_{1}, c_{1}),  (b_{2}, c_{2}), ... (b_{n}, c_{n})) $ have to be generated. As a consequence, at the end each interaction-channel contains a multitude of paths because each interaction-channel includes a split() operator.
The alternative interaction-channels that are generated by the varying application of the split() and combine() operators are processed in parallel. The effects of processing the split() and combine() operators is reflected in extensions of and changes in the interaction object. 

The rules governing the computation of the amplitudes of a path of an interaction object must be in accordance with QFT. These rules are well known and described in many textbooks
on QFT
(see for example, \cite{Ryder}, \cite{Griffiths},  \cite{Mandl}). 
However,
with QFT the respective rules are defined in terms of (external and internal ) lines and vertices of Feynman diagrams. For the functional model, the QFT rules have been mapped to rules regarding the split() and combine() operators. 


\subsubsection{Generation of "Out" Particles/Waves}

The processing of the interaction channels ends with a certain "out" particle/wave combination. With some types of interactions, different "out" particle/wave combinations may occur. The functional model of QT interactions assumes that from the possibly multiple alternative "out" particle/wave combinations only one will actually leave an interaction. 

After processing the individual interaction channels (in parallel), each interaction
channel contains the same set of paths, however, with different probability
amplitudes for the paths. Therefore, the multiple interaction channels
can be (re-)united by the summation of the corresponding amplitudes. According
to QFT rules (usually formulated in terms of Feynman diagrams), the
"summation", in some cases, has to be performed with a negative sign  (i.e., $ amplitude1 - amplitude2 $ instead of $ amplitude1 + amplitude2 $).

\subsubsection{More detailed mapping of the process steps and the system state to QFTCA}

In the above sections a relatively rough description of the CA-based functional model of QFT interactions is given.
Disregarding the size limitations for the present paper, additional, more detailed specifications would be desirable for the mapping of (1) the interaction process steps to CA-update functions and (2) the related system state to CA cells. As the simplest and initial solution it can be assumed that the whole interaction (with the above described four process steps) is performed during a single invocation of the update function and that, except for the elimination of non-interacting paths; this process affects only a single cell until the interaction results are generated.
As a next level of refinement, it is imaginable that parts of the interaction process, such as the split() and combine() operators, can be mapped to the CA-based model of waves and fields, as described in sections 4 and 5. 
This author’s computer model (see \cite{QFTCA}  has the objective to support the development of further model 
refinements.

\subsubsection{Overall Properties of QFT Interactions}

With the above described functional
model of QFT interactions, the function "QFT-interaction()" does not
constitute a bijective mapping. Step 0, "Occurrence of interaction", does not
provide a surjective mapping; only one of the multiple "in" particle positions is
mapped to the interaction result. Step 3, "Generation of "out" particles/waves",
does not represent an injective mapping; multiple "out" paths with non-zero probability
amplitudes are generated such that the ingoing state could only partly be
reconstructed from the "out" state. Altogether, the interaction process implies a
non-bijective mapping.
\footnote{
This appears to be in conflict with the QFT statement that the S-matrix is a unitary matrix and therefore an interaction is a unitary function. See also section 6.6. "The  linearity/non-linearity of QFTCA".}
As described in section 6.5, this non-bijective mapping associated with QFT interactions offers an explanation for part of the limitations and peculiarities of QT measurement.

\section{Discussions}

\subsection{How complete is the Lagrangian for the formulation of a physics theory}

The Lagrangian (and related constructs such as the Hamiltonian and the equation
of motion) is an impressively powerful basis for the formulation of theories of
areas of physics. Many details of existing physics theories can be derived from
the Lagrangian. Many past findings in physics theories have been developed by
considerations of the Lagrangian (mostly symmetry considerations). Nevertheless,
the Lagrangians do not completely specify physics theories. As shown in this
paper, mainly two additional aspects are required to enable the construction
of a comprehensive CA-based functional model that reflects subject physics
theory.
\begin{enumerate}
\item Initial states for the execution of the equation of motion
\\
The Lagrangian and equations of motion are typically differential equations.
As a trivial matter of mathematics, to obtain solutions for the
differential equations or a model that shows the dynamical evolution of a
system, some initial state (i.e., initial values for the differential equation) has
to be assumed. As shown with the sample areas of physics (sections 3, 4 and
5), however, it is not sufficient to select arbitrary initial states. Typically,
various types of constraints exist for the initial state. Only part of these
constraints can be derived from the Lagrangian. Further constraints have to
be derived from the full physics theory as described in related textbooks.
\item Process models of physics theories
\\
As described in section 1, a cellular automaton may be viewed as a specific type of a "functional model" (or "process model"), i.e., a model that
specifies the dynamical evolution of the system in terms of state transitions.
With classical areas of physics, such as Newtonian mechanics (see section 3.1),
such a process model in the form of a CA can be derived from the Lagrangian.
With more advanced physics theories, additional statements are required
for the specification of the dynamical evolution of a system. Partly, these
additional statements can be found in textbooks of the respective physics
theories. Partly, the required additional statements are still missing or still
debated among physicists. The major subject in which the provision of a process
model (here, in the form of a CA) requires additional specifications is the
area of interactions in quantum field theory (section 5).  
\end{enumerate}
 
\subsection{The role of non-local information}

The dynamical evolution of the classical CA depends exclusively on information
that can be assigned to the CA cells, and the update-function for a cell depends
only on the contents of the neighbor cells (including the respective cell). This
cell-related information is here called "local information". Differing from classical
CAs, the system state of QFTCA described in this paper contains, in addition
to the cell-related (local) information, non-local information pertaining to the
physical objects (e.g., fields and particles). While originally the inclusion of the
non-local information was merely justified by the goal to obtain a decent
model, it turned out that the inclusion of non-local information to represent
physical objects spanning larger areas of cell-space is key for the modeling of
non-local effects in connection with QFT interactions (see section 5). Non-local
effects, such as the collapse of the wave function and entanglement, can best be
explained by the assumption of physical objects with non-local attributes and
actions (see below, section 6.5 Measurement).

\subsection{The special role of time in the CA-based model}

Deviating from the trend in physics theories to treat space dimensions and the
time dimension, as much as possible, on equal footing, the Lagrangian-based CA
described in this paper assumes a special role of time with several aspects:  
\begin{enumerate}
\item Time is \emph{not} a fourth dimension of the CA,
\\
Instead, QFTCA maintains only a single time slice (i.e., the current time slice).
As described in section 2, this implies that the time derivatives appearing
in the equation of motion  (e.g., $ d \psi/dt $, $ d^{2}\psi/dt^{2} $ ) have to be kept explicitly as part of the system state. While this may be viewed as just a type
of implementation decision for the QFTCA, it also supports the author’s goal
to show that models for the evolution of our universe that assume only a
single time layer are at least feasible.
\item The existence of a time arrow
\\
The goal of providing a functional model, such as a CA-based model, of a
physics theory may be viewed as presupposing causality and an arrow of
time. Insofar, it should not be a surprise that the QFTCA described in this
paper implies an arrow of time. The neutrality with respect to a direction
of time that is represented in the (declarative) equations of physics theories,
such as the Lagrangians, is necessarily lost with the construction of a
functional (e.g., CA-based) model.

\end{enumerate}

\subsection{Consequences of discreteness}

One of the key properties of CA-based models is discreteness with respect to
space and time. Discreteness with respect to space and time coordinates may
cause discreteness with respect to further aspects, such as, for example, momentum.
Two questions arise: (1) whether discreteness impedes the precise modeling
of physics theories and (2) to which extent discreteness enables new solutions
in areas that are not yet completely understood.

\subsubsection{Does discreteness impede the precise modeling of physics theories?}
To achieve compatibility with the predictions of the standard formulations of
physics theories, the granularity of the space and time coordinates has to be
sufficiently fine. It can be expected that the finer the granularity is, the better
the agreement of the results of the CA-based modeling is with the results predicted
by the theories and the results measured in experiments. It can also be
expected (and has been confirmed by the author’s computer simulations) that
an insufficiently fine granularity may, in some cases, result in chaotic (i.e., unpredictable)
behavior of the CA. A conclusive answer to the above question requires
more complete computer modeling from the author.

\subsubsection{Does discreteness enable new solutions?}
Discreteness may result in non-linear and even unpredictable behavior (especially if there is insufficiently fine granularity). This leads to the question of whether the discreteness of the CA
may provide an explanation for the non-determinism of QT. The author’s work in
this area (including the computer model QFTCA, see \cite{QFTCA}) has so far not resulted
in a final answer to this question. While it is easy to generate unpredictability, it
has been (so far) not possible to generate through discreteness unpredictability that
adheres to the laws of QT.

As a further application of discreteness, the functional model of QFT 
(see \cite{Dielfi}), which is also the basis for the CA-based model described in section 5, assumes
discreteness with the paths of a particle. This enables a functional model of QFT
interactions that offers a new solution for the non-linearity associated with the
collapse of the wave function (see section 5) and QT measurement 
(see \cite{dielmeas} and section 6.5 below).

\subsection{QT measurement}

In \cite{dielmeas}, a functional model of QT measurement is described with the following key characteristics:
\begin{enumerate}
\item A measurement always implies interactions between the measured quantum object and the measurement environment.
\\
Measurements of QT observables can be performed using a variety of measurement
devices, apparatuses, and processes. All such measurement processes
have to include at least one interaction in which the measured object exchanges
information with some other entity belonging to the measurement apparatus.
\item The model of measurement interactions is based on QFT interactions, as described in section 5.
\\
The model assumes that the interactions between the measured QT object
and the measurement apparatus are "normal" interactions that adhere to
the laws of quantum field theory.
\item In general, the interacting particles/waves consist of multiple "paths" with different associated probability amplitudes. The interaction always occurs
at a definite position. Only the paths that cover the interaction position
determine the result of the (measurement) interaction.
\item The measurement process includes a collapse of the wave function. 
\\
After the interacting paths are determined and used to generate the interaction result (i.e., measurement result), all of the remaining paths are discarded.
This may be considered as the "collapse of the wave function".
\item As described in section 5, QFT interactions support only a non-bijective mapping
of the "in" state to the "out" state and thus only the limited exchange of
information. This limited exchange of information is the cause of some of
the limitations and peculiarities of QT measurements.
\end{enumerate}
The QFTCA described in the present paper (section 5) supports the functional model of QT measurement described in \cite{dielmeas} 
and adds further details to item (3)
(definite interaction position) and item (4) (collapse of the wave function).

\subsection{The  linearity/non-linearity of QFTCA}
Linearity is a very important feature of QT. It is therefore reasonable to discuss the linearity/non-linearity of CAs in general and of QFTCA specifically. In  \cite{Elze1}, \cite{Elze2} and \cite{Elze3},  H-T. Elze discusses the general linearity/non-linearity of CAs in detail. As a supplement to the findings described in \cite{Elze1}, \cite{Elze2} and \cite{Elze3}, the findings of the author using the computer model of QFTCA (see \cite{QFTCA}) show that an insufficiently fine granularity of the discreteness of a CA may cause a theoretically linear process to become non-linear or even chaotic.
For QFTCA specifically, the preservation of linearity is not considered an issue. QFTCA is seen as a type of "implementation" of an underlying theory (primarily QFT). Where the underlying theory requires linearity, QFTCA must provide linearity.
There are, however, two areas where QFT, the theory underlying QFTCA, does not define sufficient detail for a CA implementation; therefore, the solution(s) chosen by QFTCA may potentially violate linearity. The two aforementioned areas are the following:
\begin{enumerate}
\item The collapse of the wave function
\\
The request for the linearity of QT applies to the normal progression of the quantum system as described by the laws of QT. However, the laws of QT (in particular the equations of motion, e.g., the Schr\"odinger equation) describe only the evolution of probabilities and probability amplitudes. For the transition of the probabilities to facts (i.e., the measurement), there does
not yet exist a generally agreed-upon theory. Ideas and proposals for a theory on measurement often assume the collapse of the wave function. 
\footnote{Theories which avoid a collapse of the wave function, such as the many-worlds theory, nevertheless 
assume other types of discontinuity.}
Theories of the collapse of the wave function have a tendency towards non-linear solutions (see, for example, \cite{GWT} and \cite{Diosi}).  
QFTCA is based on the functional model of QT interactions described in \cite{Dielfi}, which also assumes a collapse of the wave function.
Differing from the theories described in  \cite{GWT} and \cite{Diosi}, which achieve non-linearity by modifications of the differential equations of the equation of motion, the functional model of QT interactions, as represented by QFTCA, utilizes the power and flexibility of functional descriptions to achieve the desired discontinuity associated with the collapse of the wave function.
\item  QFT interactions
\\
As described in section 6.5., QFTCA assumes that normal QFT interactions may represent a measurement and that, therefore, normal QFT interactions may contain a collapse of the wave function. Thus, to which extent QFT interactions may represent non-linearities (in case they include a collapse of the wave function) must be discussed.

In QFT, an important construct for the treatment of QFT interactions is the scattering matrix (S-matrix). 
In \cite{Weinberg}, Weinberg states that the S-matrix has been proven to be a unitary matrix. This characteristic of the S-matrix is often seen as a proof that QFT interactions obey linearity. However, the scattering matrix is not the equation of motion of QFT interactions. As described in sections 5, the QFT interaction does not perform a bijective mapping of the "in" state to the "out"  state. 
Whether or not this may still be considered to be a linear progression of the equation of motion is not clear to the author.
\end{enumerate} 
To summarize, it is not clear whether the above-mentioned potential non-linearities indeed represent non-linearities, and if they do, whether these non-linearities are already present within the underlying QFT is also unclear. H-T. Elze showed in  \cite{Elze3} that modifications in the underlying equation of motions may result in non-linearities. QFTCA does not assume a modification of the underlying equations of motion (or Lagrangian or Hamiltonian) but adds some interpretation details to enable the generation of a functional model.
 
\subsection{Entanglement}

The functional model of QT measurement described in \cite{Dielfi} includes a model on entanglement. The basic assumption is that the paths that represent alternative
measurement outcomes, even with normal QFT interactions (see section
5.3), encompass both correlated particles/waves. When a measurement (i.e., a
QFT interaction) selects a single path for the interacting particle/wave, the
selection also determines a specific correlation. This mechanism is not further
addressed in the present paper.

\section{Summary and Conclusions}

This paper describes an exercise to construct models of areas of physics in terms
of a cellular automaton and to derive these models (i.e., the concrete CA), as
much as possible, from the Lagrangian of the respective physics theory. Because it
was not expected that the classical CA is sufficient to model all areas of physics,
nor that the Lagrangian alone is sufficient to derive complete functional models
of the physics theories, a further aim of the exercise was to determine where the
respective limitations are and how they can be overcome with extensions (of the
CA) and the additional information required for a complete functional model.

The required extensions to the classical CA resulted in the proposal for a
QFTCA described in section 2.2. The most significant extension is the inclusion
of non-local physics objects, i.e., objects that span larger areas of cell-space
and may overlap.

The Lagrangian and equation of motion provide a powerful basis for
the provision of a (CA-based) functional model. For the modeling of advanced
areas of physics, such as QT/QFT, however, additional specifications are
required. Partly, the additional specifications can be obtained from the subject
physics theories as described in textbooks. With QFT, even further specifications,
which cannot be derived from the standard formulation of QFT, are
required. The functional model of QFT interactions described in  \cite{Dielfi} has been taken as the basis for the CA-based model described in this paper (section 5 and
6).

In addition to the determination and circumvention of the mentioned limitations,
the construction of the Lagrangian-driven CA-based functional model led to a
number of issues (see section 6: consequences of discreteness, the role of non-local
information, the special role of time, weakening the goal of linearity) that are
probably of interest beyond the scope of the present paper. With some of these
issues, only preliminary results can be offered by the author. More conclusive
answers can probably only be developed through further computer simulations.

\section*{Acknowledgment}

Discussions with Prof. H-T. Elze very much influenced  the general direction of the work described in this paper, as well as specific areas such as the statements on the linearity/non-linearity of QFTCA described in section 6.6.

%

\appendix
\section{The Computer Model QFTCA}

The computer model QFTCA is an implementation of a cellular automaton which supports the modeling of quantum field theory (QFT). The details of the particular area of QFT (e.g., quantum electrodynamics) are specified by the provision of a Lagrangian and further information related to the Lagrangian. As the major objective, the computer model supports also the modeling of interactions in QFT such as scatterings. The modeling of QFT interactions requires, first at all, some "functional model" of QFT (i.e. a model which specifies the dynamical evolution of the system in terms of state transitions). Because such a functional model of QFT cannot directly be derived from the existing standard formulation of QFT, this functional model had to be formulated first. The functional model of QFT described in \cite{Dielfi} is used as a basis.

\subsubsection{Input specifications}
The primary input is the Lagrangian or the equation of motion of the subject theory. From parameters which appear in the equation of motion further detailed specifications are prompted for
\begin{itemize}
\item the number of space dimensions to be supported by the CA,
\item the number of particles to be supported,
\item details and initial states of the individual particles,
\item details and initial states of the fields
\item initial values for the other variables which appear in the equation of motion
\end{itemize}

\subsubsection{Result of running QFTCA}
A QFTCA run shows the dynamical evolution of the system state. 
Visualization is provided for particle positions and for the probability amplitudes associated with fields and particles.

\subsubsection{Implementation}
Starting with a Lagrangian the equation of motion is determined by use of the Euler-Lagrange equation. This involves differentiation of the Lagrangian. 
The equation of motion is the basis for the update function of the CA.
The equation of motion is typically a differential equation (often second order). In general mathematics, the computation of values $ \psi (x,t) $ for fields and particles requires finding solutions for the respective differential equations.
QFTCA includes neither a "general differential equation processor" nor a "symbolic mathematics engine". 
The transformation of the Lagrangian to the equation of motion supports only a small variety of typical Lagrangians.
The determination of the dynamical evolution of state variables such as $ \psi (x,t) $ does not determine solutions of the
differential equations, but interprets the equation of motion in terms of the discrete state variables.

\end{document}